\documentclass[preprintnumbers, floatfix, preprintnumbers, letterpaper, superscriptaddress,nofootinbib, twocolumn]{revtex4}
\pdfoutput=1
\usepackage{graphicx}
\usepackage{microtype}
\usepackage{amsmath}
\usepackage{amssymb}
\usepackage{subfigure}
\usepackage{hyperref}
\usepackage{url}
\usepackage{xcolor}
\usepackage{color}
\usepackage{mathrsfs}
\usepackage{calrsfs}
\usepackage{amsfonts}
\usepackage{latexsym}
\usepackage{ragged2e}
\usepackage{epsfig}
\usepackage{textcomp}
\usepackage{phaistos}
\makeatletter
\renewcommand\@makefnmark{\hbox{\@textsuperscript{\normalfont\color{purple}\@thefnmark}}}
\renewcommand\@makefntext[1]{%
  \parindent 1em\noindent
            \hb@xt@1.8em{%
                \hss\@textsuperscript{\normalfont\@thefnmark}}#1}
\makeatother

\usepackage{caption}
\DeclareCaptionJustification{justified}{\leftskip=0pt \rightskip=0pt \parfillskip=0pt plus 1fil}
\definecolor{vividviolet}{rgb}{0.62, 0.0, 1.0}
\definecolor{amaranth}{rgb}{0.9, 0.17, 0.31}
\definecolor{palatinateblue}{rgb}{0.15, 0.23, 0.89}
\definecolor{brightpink}{rgb}{1.0, 0.0, 0.5}
\definecolor{cornflowerblue}{rgb}{0.39, 0.58, 0.93}
\definecolor{deepcarminepink}{rgb}{0.94, 0.19, 0.22}
\definecolor{radicalred}{rgb}{1.0, 0.21, 0.37}

\hypersetup{ linktoc=all,
    colorlinks, linkcolor={palatinateblue},
    citecolor={brightpink}, urlcolor={amaranth}
}

\graphicspath{{Images/}}

\renewcommand{\d}[1]{\ensuremath{\operatorname{d}\!{#1}}}

\graphicspath{{Images/}}
\renewcommand{\d}[1]{\ensuremath{\operatorname{d}\!{#1}}}
\makeatletter
\def\@fnsymbol#1{\ensuremath{\ifcase#1\or $\textleaf$ \or $\PHplaneTree$
\else\@ctrerr\fi}}%

\makeatother

\def\sideremark#1{\ifvmode\leavevmode\fi\vadjust{\vbox to0pt{\vss
 \hbox to 0pt{\hskip\hsize\hskip1em
 \vbox{\hsize1.5cm\tiny\raggedright\pretolerance10000
 \noindent #1\hfill}\hss}\vbox to8pt{\vfil}\vss}}}%
\begin{document}

\title{The Charge of Electron, Weak Gravity Conjecture and Black Hole Evolution}

\author{Yen Chin \surname{Ong}}
\email{ycong@yzu.edu.cn}
\affiliation{Center for Gravitation and Cosmology, College of Physical Science and Technology, Yangzhou University, \\180 Siwangting Road, Yangzhou City, Jiangsu Province  225002, China}
\affiliation{School of Aeronautics and Astronautics, Shanghai Jiao Tong University, Shanghai 200240, China}

\begin{abstract}
The charge of an electron is vastly larger than its mass. We found that black hole evolution under Hawking evaporation is very sensitive to the value of electron charge.
If the electron charge is weakened by a mere one order of magnitude, then the evolutionary paths of Reissner-Nordstr\"om black holes under Hawking evaporation change substantially. In particular, weakening the electron charge causes some black holes that would otherwise immediately discharge towards Schwarzschild limit to first evolve towards extremality before turning around. We discuss the possible connections between the weak gravity conjecture, the cosmic censorship conjecture, and black hole evolution.
\begin{center}

\end{center}
\end{abstract}

\maketitle

\section{Introduction: Gravity is Weak}\label{1}

In elementary physics textbooks and popular science books it is often written that gravity is the weakest force, because the ratio of the gravitational force
\begin{equation}
F_\text{G}=\frac{GMm}{r^2},
\end{equation}
to that of the electromagnetic force
\begin{equation}
F_\text{EM}=\frac{1}{4\pi \epsilon_0}\frac{Qq}{r^2},
\end{equation}
between two electrons is small: $F_\text{G}/F_\text{EM} = \mathcal{O}(10^{-42})$. 
See, e.g., p.6 of \cite{bruce}, in which after mentioning this fact, wrote that ``gravity is intrinsically weak''.
Such statement, however, is not quite a precise characterization of the weakness of gravity, since a hypothetical particle with Planck mass $m_\text{Pl}:=\sqrt{\hbar c/G}$ and Planck charge $q_\text{Pl}:=\sqrt{4\pi\epsilon_0\hbar c}$ would have $F_\text{G}/F_\text{EM}=1$. 

The real mystery is this: why is the charge of electron so much higher than its mass? Of course, charge and mass have different physical dimensions so they cannot be directly compared. The precise statement is that the \emph{dimensionless ratio} $m_e/m_\text{Pl}= \mathcal{O}(10^{-22}) \ll q_e/q_\text{Pl} = \mathcal{O}(10^{-1})$. From now onwards we will use the ``relativistic units'' \cite{werner} in which $c=G=4\pi \epsilon_0=1$ (but $\hbar \neq 1$) so that charge and mass both have the same dimension of length and so we can directly write $m_e \ll q_e$. The charge-to-mass ratio $q_e/m_e$ is about $10^{21}$ in the relativistic units. 

This observation -- along with the fact that gravity is also significantly weaker when compared to the strong and weak nuclear forces --  is known as the \emph{hierarchy problem}, and has led to numerous theoretical investigations. 

The ``weak gravity conjecture'' (WGC) was then proposed \cite{0601001}, which essentially states that the lightest charge particle with mass $m$ and charge $q$ in any $U(1)$ gauge theory that admits an ultraviolet embedding into a consistent theory of quantum gravity should satisfy the nontrivial bound
\begin{equation}
\frac{gq}{\sqrt{\hbar}} \gtrsim \frac{m}{m_\text{Pl}},
\end{equation}
where $g$ is the coupling constant of the $U(1)$ force. 
The electron satisfies the WGC. In fact, since $g=q_e/\sqrt{\hbar}$ for electron-photon coupling\footnote{This is just the square root of the fine structure constant, i.e. $g=\sqrt{\alpha}=q_e/q_\text{Pl}$.}, the inequality is satisfied by a \emph{huge} margin in quantum electrodynamics (QED):
\begin{equation}
\frac{gq_e}{\sqrt{\hbar}} \sim 10^{-3} \gg \frac{m_e}{m_\text{Pl}} \sim 10^{-22}.
\end{equation}
The statement that gravity is weak compared to electromagnetism can be rephrased as: why is WGC satisfied by such a large margin by the electron? 

The WGC ensures that, among other things, an extremal black hole is unstable and can thus decay into non-extremal black holes via charged particle production (despite its Hawking temperature being zero) \cite{0601001, 1611.06270}. One might even suspect that with the WGC, near-extremal black holes can discharge efficiently and avoid becoming extremal in the first place. If so, WGC also implies the third law of black hole thermodynamics (that zero temperature state cannot be reached in finite steps -- here emission of a Hawking quanta is a step). In fact, this is exactly the case for an asymptotically flat Reissner-Nordstr\"om black hole. 

In an interesting work \cite{HW}, Hiscock and Weems showed that under Hawking evaporation, if the charge-to-mass ratio is initially small, then it can in fact \emph{increase} as the (isolated) black hole evaporates. The ratio can even approach extremality but never attains it: at some point the evolution turns around towards the Schwarzschild limit instead. If the initial charge-to-mass ratio is sufficiently high, than the black hole simply discharges towards the Schwarzschild limit. See Fig.(\ref{HWfig}) below. This is satisfactory because it demonstrates nicely how the third law is satisfied, albeit it comes very close to being violated. (The fact that extremality is never reached also means that the cosmic censorship conjecture \cite{CCC} holds; see Sec.(\ref{3}) for discussion.) We will review the result of Hiscock and Weems in Sec.(\ref{2}).

Since there is a huge margin for QED to satisfy the WGC, one would naturally expect that how evaporating charged black holes evolve should be rather insensitive to the ratio $q_e/m_e$, since the black hole charge and mass are a lot larger than that of electrons. However, as we shall show below, surprisingly this is not the case, and by lowering $q_e/m_e$ by an order of magnitude changes the picture. Even though the third law is never violated, a relatively small change (as small as one order of magnitude) in $q_e/m_e$ leads to very different evolutionary history of charged black holes. That is, \emph{black hole evolution is surprisingly sensitive to the value of electron charge}.

Note that, as we commented in Footnote 1, in our units $\alpha=q_e^2/\hbar$, so with fixed $\hbar$, changing the value of $q_e$ is the same as varying the fine structure constant. Thus our work can also be interpreted as the study of the effect of varying $\alpha$ on charged black hole evaporation. Of course in a realistic universe, changing $\alpha$ would also affect stellar physics, which would in turn affect black hole formation. However our aim is more modest: \emph{given} a black hole, which is still a valid solution to the Einstein field equations, how would it evolve?\footnote{The effect of varying $\alpha$ on other aspects of black hole physics was investigated in \cite{0212334}.}

\section{How Charged Black Holes Evolved in Einstein-Maxwell Theory}\label{2}

Following Hiscock and Weems, we use the units in which $c=G=k_B=4\pi \epsilon_0=1$, while $\hbar=\hbar G/c^3 \approx 2.61 \times 10^{-66} \text{cm}^2$. Mass and charge both have dimension of length (since, e.g. $M=GM/c^2$). In the following their units will be in centimeters.
A solar mass is about 1.5km.

Reissner-Nordstr\"om black hole is the solution to the Einstein-Maxwell theory. Its metric tensor is
\begin{flalign}
g[\text{RN}]=&-\left(1-\frac{2M}{r}+\frac{Q^2}{r^2}\right)\d t^2\\ \notag & + \left(1-\frac{2M}{r}+\frac{Q^2}{r^2}\right)^{-1}\d r^2 + r^2\d\Omega^2_{S^2},
\end{flalign}
with the event horizon at $r_+=M+\sqrt{M^2-Q^2}$. The case with $M=Q$ is the extremal black hole. Without loss of generality and for convenience of notation, we follow \cite{HW} and set $Q>0$. Due to like charges repel, the black hole predominantly emits positively charged particles. That is, technically we will be discussing positron, with charge that we will denote as $e:=|q_e|$. Of course our results hold for both electrons and positrons, \emph{mutatis mutandis}. The numerical value of the charge is $e=6 \times 10^{-34}$cm, while the mass is $m:=m_e\approx 10^{-21} e$. 

The Hawking temperature of the black hole is
\begin{equation}
T=\frac{\hbar\sqrt{M^2-Q^2}}{2\pi \left(M+\sqrt{M^2-Q^2}\right)^2},
\end{equation}
and we see that in the $M \to Q$ limit the extremal black hole has zero temperature. This does not mean the black hole ceases to radiate. In fact, Hawking calculated that the number of particles of the $j$-th species
with charge $q$ emitted in a wave mode labeled by frequency $\omega$, spheroidal harmonic $l$, and
helicity $p$ is given by (in the case of static black hole and zero angular momentum particles) \cite{Hawking2}, 
\begin{equation}\label{Hawking}
\left\langle N_{j\omega l p} \right\rangle = \frac{ \Gamma_{j\omega l p}}{\exp\left[(\omega-e\Phi)/T\right]\pm 1}.
\end{equation}
The plus and minus signs in the denominator correspond to fermion and boson, respectively. $ \Gamma_{j\omega l p}$ denotes the absorption probability for an incoming wave of the specific mode. One notes that for a sufficiently strong electric field, particle number need not be zero even if $T=0$. 

Since the temperature is inversely proportional to the black hole size, a sufficiently large black hole will also be cold enough that the creation of massive particles is suppressed exponentially by the Boltzmann factor.  In this regime, as shown by Gibbons \cite{gibbons}, as well as Hiscock and Weems, the thermal channel will emit only massless particles, while the exponentially suppressed charge loss can be modeled via the Schwinger formula. 
That is, the evolution under particle production
 is governed by two coupled ordinary differential equations:
\begin{equation}\label{dMdt}
\frac{\d M}{\d t} = -a \beta \sigma T^4 + \frac{Q}{r_+}\frac{\d Q}{\d t},
\end{equation}
\begin{equation}\label{dQdt}
\frac{\d Q}{\d t} \approx -\frac{e^4}{2\pi^3\hbar m^2}\frac{Q^3}{r_+^3}\exp\left(-\frac{r_+^2}{Q_0Q}\right).
\end{equation}
Note that under the Hawking process, charge particle production is only suppressed, not entirely absent.  The Hiscock-Weems model captures this phenomenon, albeit with non-thermal charge loss\footnote{``\emph{All models are wrong, some are useful.}'' -- George E. P. Box.}. 
Thus, as emphasized by Hiscock and Weems, both $\d M/\d t$ and $\d Q/ \d t$ terms above are part of the Hawking radiation\footnote{In the literature, it is common to refer to only the thermal channel as ``Hawking radiation''. One can then view the Hiscock and Weems model as describing the evolution of charged black holes under both Hawking radiation and the Schwinger effect. These separations are largely semantic as far as our purpose is concerned. Note that Hiscock and Weems model works also for extremal black holes: $\d Q/\d t$ and hence $\d M/ \d t$ is nonzero although $T=0$. This decay is allowed assuming weak gravity conjecture holds.}. 

The first term on the RHS of Eq.(\ref{dMdt}) is the Stefan-Boltzmann law, where $a$ denotes the radiation constant $a=\pi^2/(15\hbar^3)$, and $\sigma$ the effective emission area, which in the geometric optics approximation is equal to the area of a sphere whose radius is the impact parameter. In the case of Reissner-Nordstr\"om black hole, we have \cite{HW}
\begin{equation}
\sigma[\text{RN}]=\frac{\pi}{8}\frac{(3M+\sqrt{9M^2-8Q^2})^4}{(3M^2-2Q^2+M\sqrt{9M^2-8Q^2})}.
\end{equation}
Due to scattering at long wavelengths, the effective emission surface is actually smaller than the one given by the geomertic optics cross section. This is governed by $\beta$, the greybody factor in Eq.(\ref{dMdt}). However, since its effect is tiny compared to the black hole lifetime (see \cite{HW}), we will set it to be unity. Whereas for the $\d Q/\d t$ expression, Eq.(\ref{dQdt}), $Q_0$ is the inverse of the Schwinger critical field: $E_c:={m^2 c^3}/{e\hbar} = 1.312 \times 10^{16}~\text{V/cm}$. The Hiscock-Weems model works for sufficiently large black holes $M \gg Q_o := \hbar e/(\pi m^2)$ \cite{HW}. The black hole lifetime is extremely long, so one can view the process as quasi-static.
\begin{figure}[!h]
\centering
\includegraphics[width=3.2in]{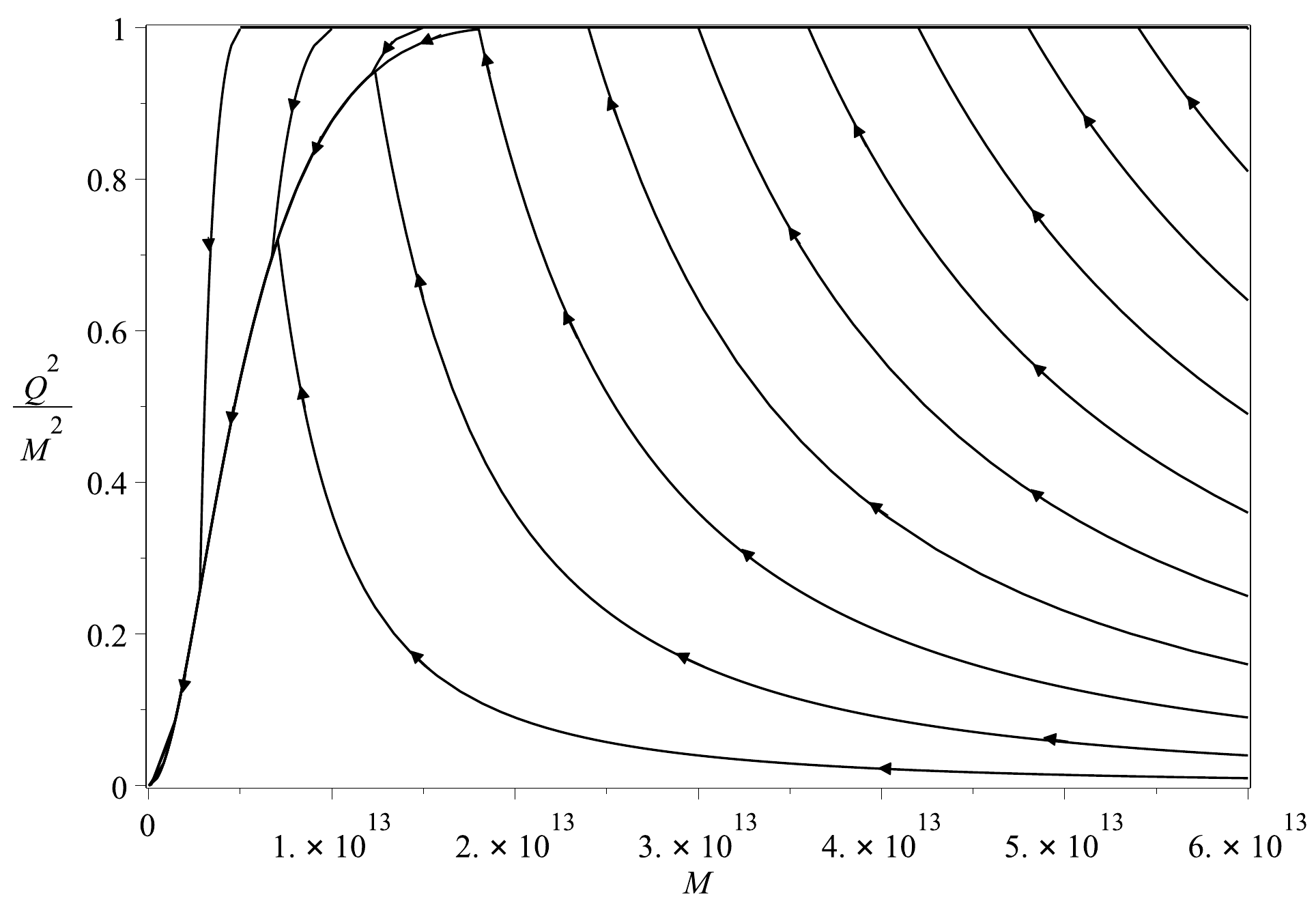}
\caption{The evolution of asymptotically flat Reissner-Nordstr\"om black holes under Hawking evaporation with Hiscock-Weems model. Arrows denote the direction of evolution. Note that there exists an attractor, which evolves towards the Schwarzschild limit, while it may look like some curves at larger values of $M$ reaches $(Q/M)^2=1$, what happened is that they eventually hit the attractor curve (which tends to unity at large $M$) and slowly discharge towards the Schwarzschild limit \cite{HW}.
\label{HWfig}}
\end{figure}

We can now numerically investigate how black holes evolve under the coupled ODEs Eq.(\ref{dMdt}) and Eq.(\ref{dQdt}). The result is given in Fig.(\ref{HWfig}), which can also be found in \cite{HW}. 
The main feature is the existence of an attractor curve. Once other curves hits the attractor curve\footnote{The word ``hit'' is used colloquially here. Although the curves come very close together near the attractor, none of the curves ever intersect, since the evolution of ODE is unique.} they will evolve along it and flows to the Schwarzschild limit ($Q\to 0$), thus extremality cannot be reached, although some black holes come very close to being extremal. The third law of black hole thermodynamics is thus satisfied under Hawking evaporation in Einstein-Maxwell theory.

\section{Black Holes Evolution with Different Electron Charge}\label{3}

\begin{figure}[!h]
\centering
\includegraphics[width=3.2in]{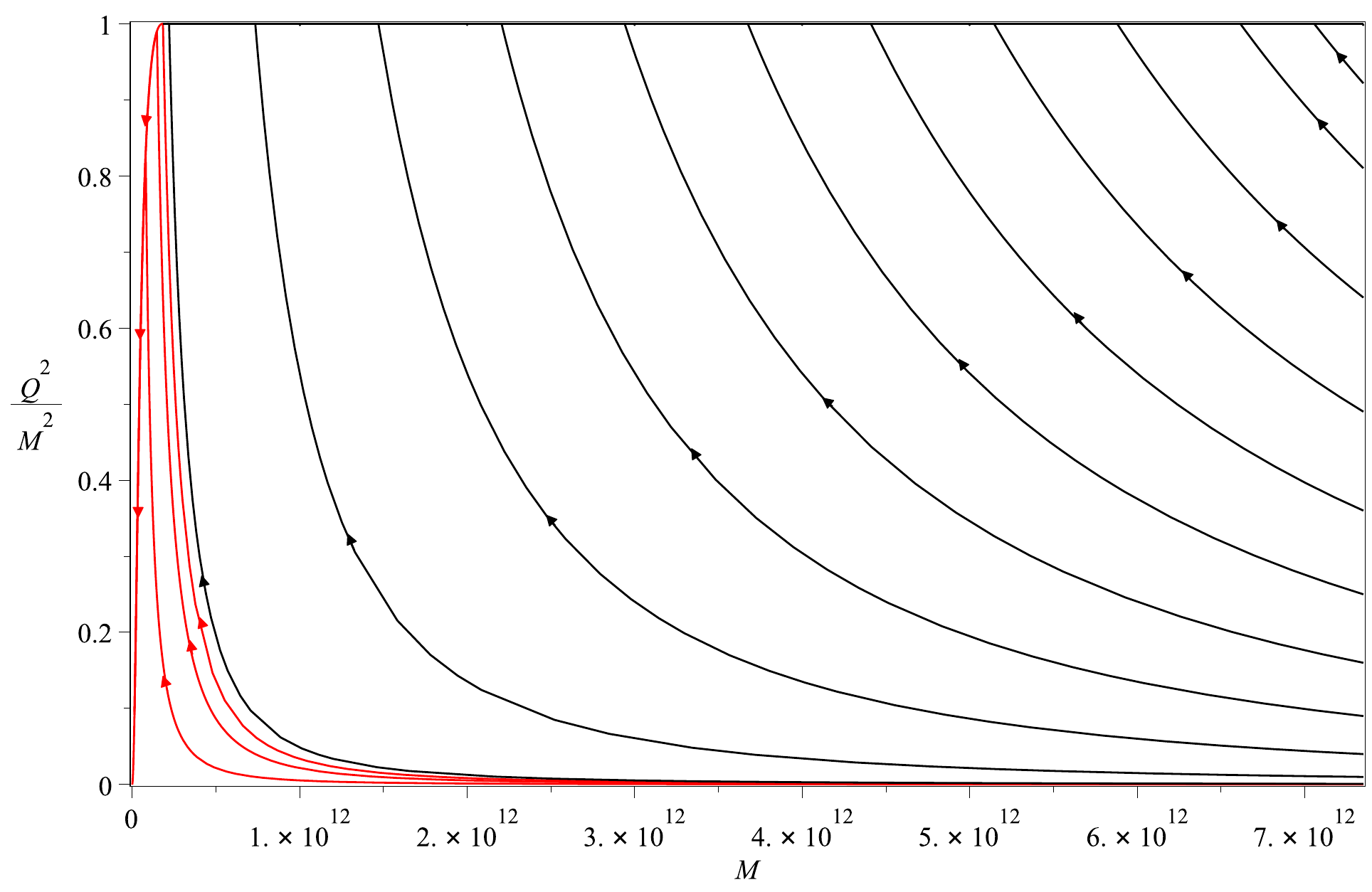}
\caption{The evolution of asymptotically flat Reissner-Nordstr\"om black holes under Hawking evaporation with Hiscock-Weems model, but with $e/m=10^{19}$ instead of $10^{21}$. Arrows denote the direction of evolution. The position of the attractor has shifted towards the left part of the plot. The lowest 3 curves in red show trajectories that hit the part of the attractor that starts to flow downward towards Schwarzschild limit, at around  $M=1\times 10^{12}$cm. 
\label{em2}}
\end{figure}

We now repeat the calculation but with a different ratio of $e/m$. 
The resulting plot of $(Q/M)^2$ against $M$ of the black hole evolution is qualitatively similar to Fig.(\ref{HWfig}). However, the attractor has shifted. The point where the attractor starts to flow towards the Schwarzschild limit occurs at smaller value of $M$, at around $M=1\times 10^{12}$cm instead of $M=1.8 \times 10^{13}$cm.
Fig.(\ref{em2}) shows the plot of $(Q/M)^2$ versus $M$, assuming that $e/m=10^{19}$ instead of $e/m=10^{21}$. 

Decreasing the charge-to-mass ration $e/m$ further to $10^{18}$ pushes the attractor further down the mass scale: the part where the near-extremal black hole starts to discharge is of the order  $M=\mathcal{O}(10^9)$cm.
This is easy to understand: weakening the electron charge means discharge via the Schwinger formula is less efficient.
Taking initial value $M(0)=7.35  \times 10^{12}$cm and $Q(0)=0.2M(0)$, we can verify that within numerical accuracy up to 16 digits of significant, $M=Q=1.470000000000000 \times 10^{12}$ at $t \sim 10^{100}$ years, a long\footnote{This is the usual characteristic of Hawking evaporation: it is a slow process due to the temperature being extremely low. A solar mass Schwarzschild black hole takes $\mathcal{O}(10^{67})$ years to evaporate. A charged black hole, with lower temperature, would take a vastly longer time. See \cite{HW} for detailed discussions.} but finite amount of time. 
They start to differ at the \emph{17th} digit of significant. One certainly should not trust the numerical values too much near extremality (when large numbers are involved), so there is a danger that extremality can be obtained in finite time. However, since the feature of the evolution remains the same qualitatively, and the attractor appears at mass scale $M \gg Q_0$, i.e. in the regime the model can be trusted, it seems likely that extremality is not reached, and instead the black hole eventually flows down along the attractor.

\begin{figure}[!h]
\centering
\includegraphics[width=3.2in]{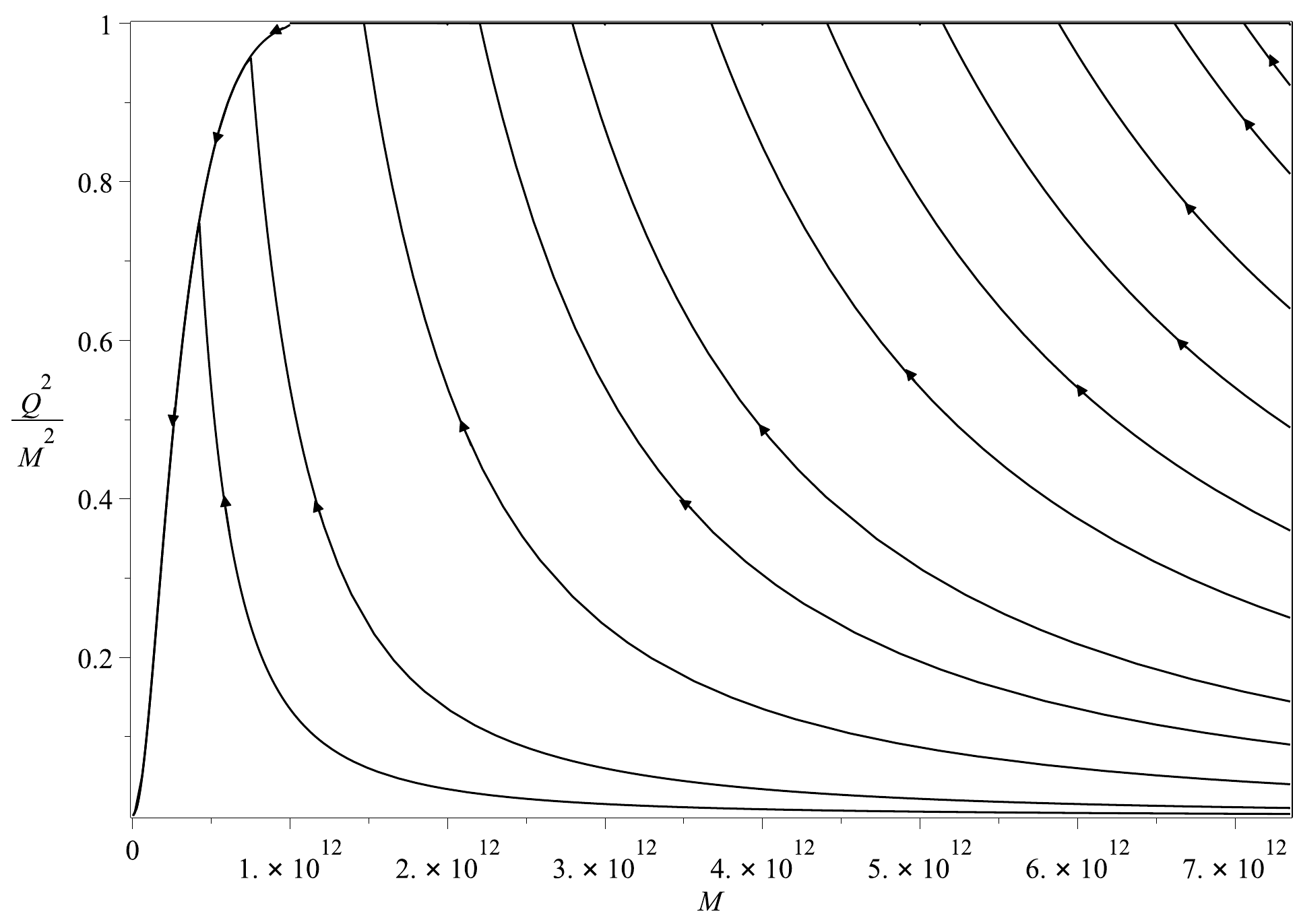}
\caption{The evolution of asymptotically flat Reissner-Nordstr\"om black holes under Hawking evaporation with Hiscock-Weems model, but with $e/m=10^{20}$ instead of $10^{21}$. Arrows denote the direction of evolution.
\label{em1}}
\end{figure}

\begin{figure}[!h]
\centering
\includegraphics[width=3.2in]{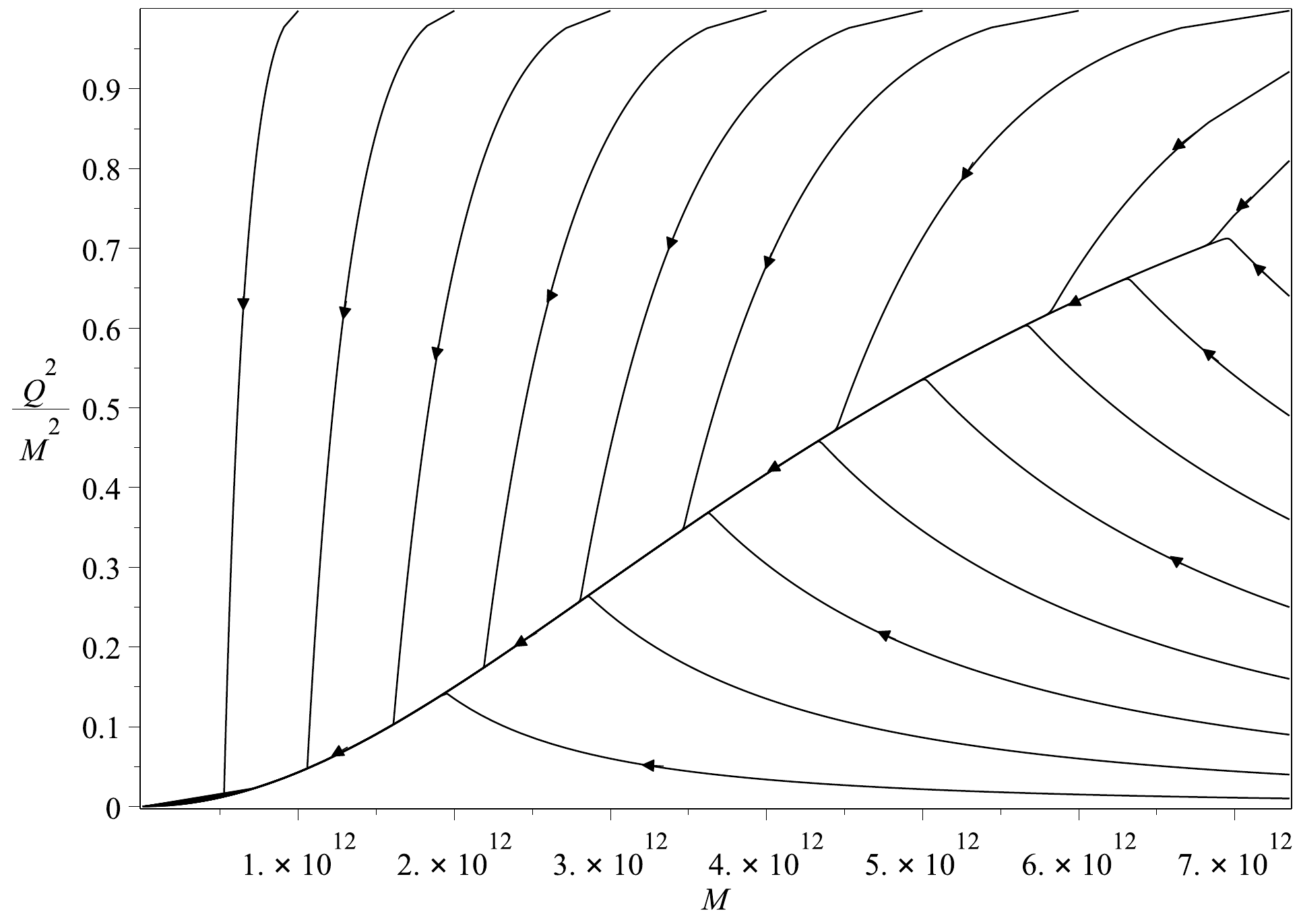}
\caption{The evolution of asymptotically flat Reissner-Nordstr\"om black holes under Hawking evaporation with Hiscock-Weems model, assuming the original ratio $e/m=10^{21}$. Arrows denote the direction of evolution. The plot is of the same mass range as Fig.(\ref{em1}) yet it is very different.
\label{em1b}}
\end{figure}

In any case, it is clear that the evolutionary paths of Reissner-Nordstr\"om black holes change quite drastically as we reduce the ratio $e/m$. This is true even if $e/m$ is only increased by \emph{one} order of magnitude. In Fig.(\ref{em1}) we show the plot of $(Q/M)^2$ versus $M$ assuming $e/m=10^{20}$. If we compare it with the plot with the original ratio of $e/m$, Fig.(\ref{HWfig}), over the same mass range, which for convenience we show in Fig.(\ref{em1b}), we see that the positions of the attractor are very different. 

Indeed, a huge portion of the curves in Fig.(\ref{em1}) is in the mass dissipation regime \cite{HW}, where black holes lose mass faster than charge, causing its charge-to-mass ratio to first increase, until it hits the attractor and turns towards the Schwarzschild limit. In contrast, in Fig.(\ref{em1b}), the mass dissipation regime is almost the same size as the charge dissipation regime, in which black holes discharge steadily.

\section{Conclusion}\label{3}

In this work, we have shown that although the electron satisfies the weak gravity conjecture by a huge margin, black hole evolution is \emph{highly sensitive} to the electron charge-to-mass ratio.
Weakening the electron charge by one order of magnitude is sufficient to cause black holes to evolve very differently. In particular, some black holes that would otherwise immediately discharge towards the Schwarzschild limit would now first increase its charge-to-mass ratio, before hitting an attractor curve. The change is actually already noticeable at smaller deviation of $e/m$, say at $e/(2m)$, though not as dramatic, unless one goes to very large values of the mass. This is surprising since the charge-to-mass ratio of a black hole involved can be much larger, i.e. $\mathcal{O}(1)$, compared to $\mathcal{O}(10^{-21})$ of the electron's charge-to-mass ratio.

In the work of Hiscock and Weems \cite{HW} it was claimed that extremality is never reached because the near-constant attractor at large $M$ is characterized by the boundary of a positive specific heat region, which itself tends to but never reaches, extremality. However, the relationship between the specific heat and the attractor requires further study because as shown in \cite{1907.07490}, the asymptotically flat dilaton charged black hole \cite{g, gm, ghs} also has an attractor, but the specific heat is always negative.
We will return to this issue in future work.

The statement that gravity is weak compared to electromagnetism can be rephrased as follows: why is WGC satisfied by such a large margin by the electron? While our work does not answer this long-standing question, it does reveal that if the electron charge (equivalently, the fine structure constant) is not its present value, then charged black holes would evolve drastically differently. 

The WGC, if correct, would allow \emph{extremal} black hole to decay via emission of charged particles. It is possible that it also prevents \emph{near extremal} black holes from reaching extremality. The Hiscock-Weems model is a concrete example of this. This ensures that the cosmic censorhsip is never violated under Hawking evaporation. Based on another non-trivial example from charged dilaton black hole \cite{1907.07490}, \emph{we conjecture that under black hole evolution due to Hawking radiation, the black hole parameters always evolve in such a way as to eventually avoid violating cosmic censorship.} Since there is evidence that the WGC is related to cosmic censorship \cite{1709.07880, 1901.11096}, it is possible that such evolution (that avoids violating the censorship conjecture) only occurs because of the weakness of gravity.

Finally, we note that, as shown in Fig.(\ref{HWfig}), the turn over towards Schwarzschild limit starts at around $M \sim 10^8$ solar masses. Considering the standard model of particle physics, and taking into considerations the greybody factor would affect this number by 1-2 order of magnitude (see \cite{HW}). Interestingly, in our actual Universe, supermassive black holes range from around $10^6$ to $10^{10}$ solar masses. Typical black holes in the galaxies are much smaller, ranging from about 5 to several tens of solar masses. Of course these black holes are \emph{not} Reissner-Nordstr\"om black holes, but rather rotating ones (furthermore, they are rarely isolated and so do not carry much electrical charge; and in addition real black holes absorb more cosmic microwave photons than emitting Hawking radiation) so our analysis does not apply. Nevertheless, it is fun to note that if we ignore rotation, this means that the value of electron charge is such that most black holes cannot approach extremality, only truly massive ones can do so and stay close to extremality for a long time. More research is needed to examine the rotating case more carefully.

\begin{acknowledgments}
YCO thanks the National Natural Science Foundation of China (No.11705162) and the Natural Science Foundation of Jiangsu Province (No.BK20170479) for funding support. He also thanks Brett McInnes for useful suggestions.
\end{acknowledgments}

\end{document}